\newcommand{\apj}{ApJ}
\newcommand{\apjl}{ApJL}
\newcommand{\mnras}{MNRAS}

\newcommand{\araa}{ARAA}

\documentclass[
    ,final            
  ]
  {aipproc}

\usepackage{amssymb}
\usepackage{amsmath}

\layoutstyle{6x9}

\begin{document}

\title{Population III Star Formation During and After the Reionization Epoch}


\author{Michele Trenti}{
  address={Department of Astrophysical \& Planetary Sciences, University of Colorado, CASA, 389-UCB, Boulder, CO 80309, USA}
}

\begin{abstract}
 
  Population III star formation during the dark ages shifted from
  minihalos ($\sim 10^6~ M_{\odot}$) cooled via molecular hydrogen to
  more massive halos ($\sim 10^8~ M_{\odot}$) cooled via Ly-$\alpha$
  as Lyman-Werner backgrounds progressively quenched molecular hydrogen 
  cooling. Eventually, both modes of primordial star formation were 
  suppressed by the chemical enrichment of the IGM.  We present a 
  comprehensive model for following the modes of Population III star
  formation that is based on a combination of analytical calculations 
  and cosmological simulations. We characterize the properties of the
  transition from metal-free star formation to the first Population 
  II clusters for an average region of the Universe and for the 
  progenitors of the Milky Way. Finally, we highlight the possibility of 
  observing the explosion of Population III stars within Ly-$\alpha$ 
  cooled halos at redshift $z \sim 6$ in future deep all sky surveys 
  such as LSST.

\end{abstract}

\maketitle


\section{Introduction}

The first luminous objects in the Universe, metal-free or Population
III stars, are thought to form in dark matter halos with virial 
temperatures $T_{vir} \gtrsim 10^3$ K corresponding to masses $M_h \sim 
10^5-10^6~M_{\odot}$, depending on redshift \cite{tegmark97}.  In these 
``minihalos'', molecular hydrogen cooling lowers the gas temperature to 
$\sim 200$ K, which leads to the formation of either a single star or a 
binary that may be very massive ($O(100~M_{\odot})$) \cite{abel02,bromm04,
oshea07,turk09}. The first stars begin forming in mihihalos very early in 
the Universe, at redshift $z\gtrsim 60$ \cite{naoz06,ts07a}, and might 
provide the first seeds of supermassive black holes if their mass is 
sufficient to directly collapse into black holes with $M_{BH} \gtrsim 
100~M_{\odot}$ \cite{heger03}. Interestingly, the locations of minihalos 
hosting the very first stars are not strongly correlated with the most 
massive clusters at $z = 0$.  Their remnants instead likely reside in 
halos with masses $M_{h} \lesssim 3 \times 10^{13} ~M_{\odot}$ today 
\cite{tss08}.

As redshift decreases, the metal-free star formation rate rapidly
increases because minihalos become progressively more abundant, until 
negative feedback by Lyman Werner (LW) photons photodissociates H$_2$
molecular and increases the minimum halo mass required for cooling
\cite{haiman97}. This happens at $z \lesssim 35$, once Population III 
star number densities $n_{PopIII}$ reach $\sim 1~\mathrm{Mpc^{-3}}$
(comoving). After that, a self-regulated phase for metal-free star
formation sets in \cite{ts09}. Eventually, the LW background becomes so
strong that only halos with $T_{vir} \gtrsim 10^4$ K are able to cool.
This likely happens at $z \lesssim 15$, when the first metal-enriched
galaxies have already been formed \cite{ricotti08,trenti09b}. 
Metal-enriched and metal-free star formation coexist in different 
regions of the Universe, with metal-enriched galaxies primarily in
overdensities and metal-free stars in voids \cite{trenti09b,stiavelli09}. 
Population III stars continue forming until $z\lesssim 5$, although at 
very low rates \cite{tornatore07,trenti09b}. The precise details of late
Population III star formation are difficult to study, however. The box
sizes needed to enclose rare metal-free star formation at low redshifts, 
comoving volumes $\gtrsim 10^3 \mathrm{Mpc^3}$, must also be able to 
resolve the DM halos \cite{bagla05,trenti10a}.  Doing this with the 
necesary sub-grid physics and/or post-processing analysis would severely 
tests the memory limits of any current model.  In addition, Population 
III star formation might proceed differently in partially ionized gas 
and possibly result in less massive stars \cite{yoshida07}. The initial 
mass function (IMF) of metal enriched stars is also uncertain at very 
high redshift because the temperature of protostellar clouds couples 
to the CMB temperature. Hence, a complex scenario with multiple modes of 
metal enriched star formation might emerge \cite{smith_b09,schneider10}.

\begin{figure}\label{fig:sfr}
 \includegraphics[height=.32\textheight]{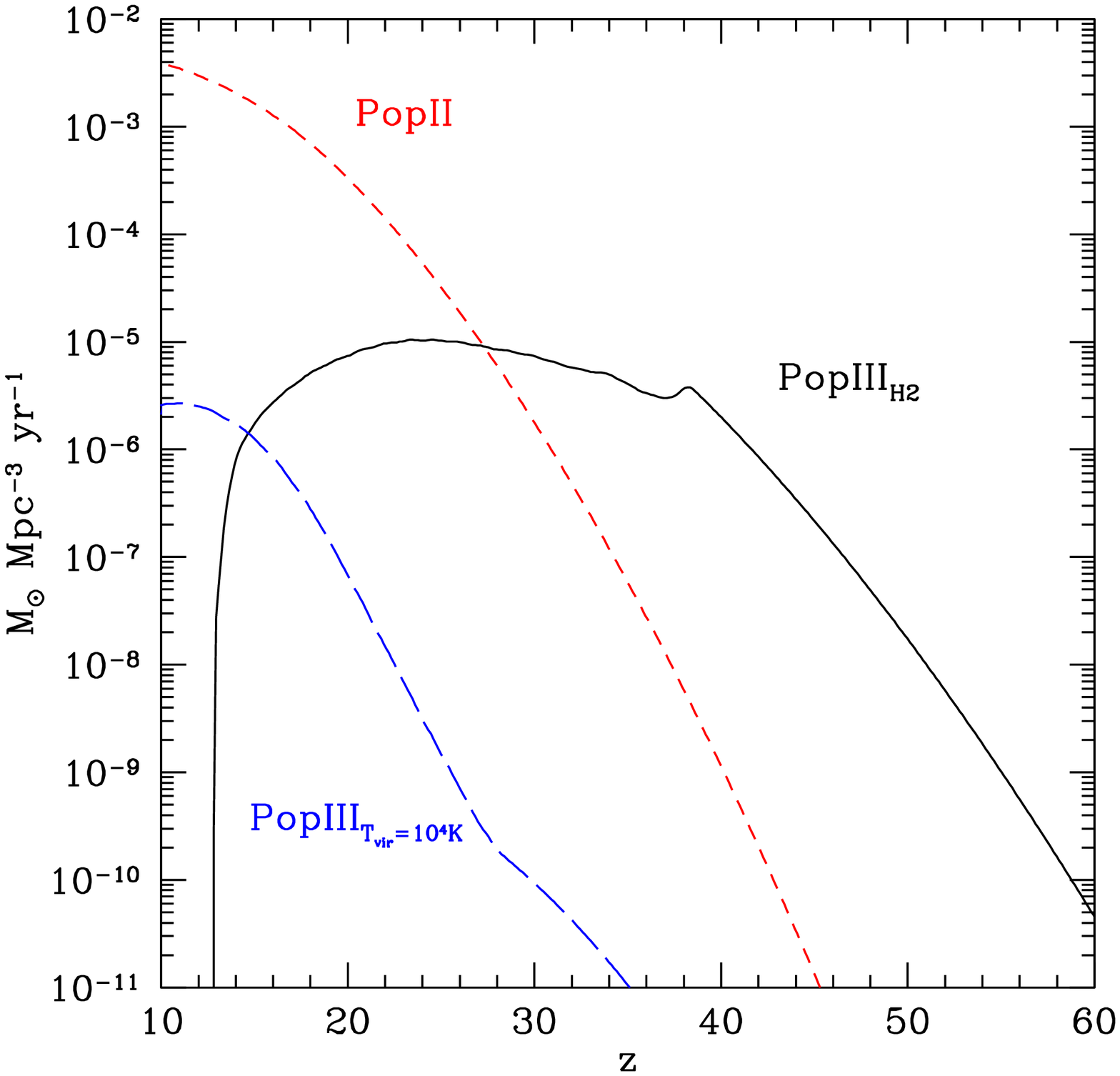}
\includegraphics[height=.32\textheight]{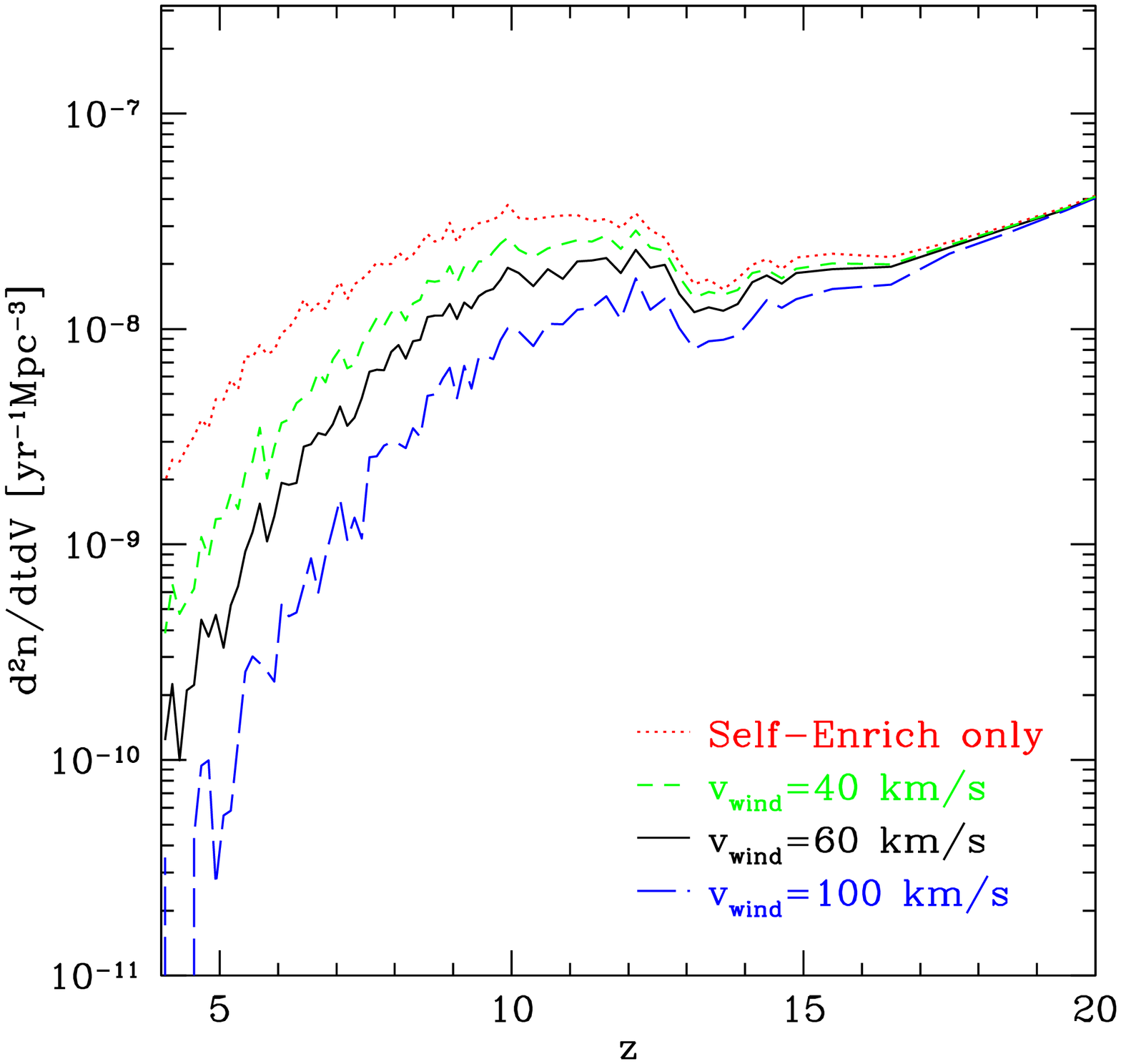}
 \caption{Left panel: star formation rates versus redshift predicted
   by our analytical model for the Population III to Population II
   transition (solid black line: Population III stars in minihalos;
   long dashed blue line: Population III stars in more massive halos,
   with $T_{vir}\geq 10^4 \mathrm{K}$; short dashed red line:
   Population II stars formed out of metal-enriched gas). Figure is from
   \cite{ts09}. Right panel: late-time Population III halo formation
   rate versus redshift derived from a cosmological simulation of
   structure formation, post-processed to include metal enrichment by
   progenitor halos and by metal outflows from neighbor halos.
   Curves from top to bottom refer to different metal outflow
   velocities. Even under the assumption of efficient metal transport
   (high $v_{wind}$), pockets of metal-free gas capable of forming
   stars exist at $z\sim 5$. Figure is from \cite{trenti09b}.}
\end{figure}

\section{Population III to Population II Transition}

To address star formation in this complex framework, we start from
dark-matter-only halo dynamics, based either on Press-Schechter
models complemented by analytical treatments of self-enrichment 
\cite{ts09} or on detailed halo-merger histories derived from 
cosmological simulations \cite{trenti09b}. We then populate halos 
with stars based on an analytical cooling model that accounts for 
redshift, metallicity and radiative feedback in the LW bands (see 
\cite{ts09} for details).

Figure~\ref{fig:sfr} shows our results: metal-free star formation is
an extended process over redshift. Only at extremely high redshift ($z
\gtrsim 35$) does Population III star formation depends just on local 
halo properties. Based on nomenclature by \cite{tan08}, these are 
Population III.1 stars. At lower redshift, metal-free star formation 
is also affected by radiative feedback, and such stars are Population 
III.2 stars. Some regions of the Universe continue to host Population 
III stars down to relatively low redshift ($z\lesssim 5$), although at 
very low rates (right panel of Figure~\ref{fig:sfr}). These late-time 
Population III stars form in halos with $T_{vir} \gtrsim 10^4$ K that 
are massive enough to cool even in presence of a strong LW background. 
The dominant process that quenches Population III star formation under 
these conditions is wind enrichment from neighbor protogalaxies. Because 
of this, late-time metal-free sources tend to be antibiased, and are 
preferentially formed in underdense regions \cite{stiavelli09}. Even if 
they are clustered, these sources are too faint for direct detection in 
the near future. A $100~M_{\odot}$ star at $z=6$ is about 7 magnitudes 
fainter than the detection limit of the James Webb Space Telescope for 
a deep field survey. However, a small cluster of Population III stars 
could be observed if magnified by gravitational lensing 
\citep{stiavelli09}. Alternatively, all-sky surveys reaching $M_{AB} 
\sim 26$ could detect Population III stars if they explode as bright 
supernovae \cite{trenti09b}.

Our model also highlights that metal-enriched star formation becomes
the dominant mode of star formation at $z \lesssim 25$ (left panel of
Figure~\ref{fig:sfr}). This implies that Population III stars play a 
minor role in the reionization of the Universe \cite{ts09}, with normal 
galaxies producing most of the ionizing photons \cite{trenti10b}.


Because direct investigation of the properties of metal-free stars 
is difficult, the properties of the most metal-poor stars in the 
local Universe have been studied to infer the IMF of their progenitors 
\cite{freeman02,beers05,frebel07,tumlinson04}. The idea behind these 
``galactic archeology'' campaigns is that extremely metal-poor stars are 
likely to be objects formed out of gas enriched only by one previous 
generation of stars. If the first generation of stars is very massive 
($M \gtrsim 100 ~M_{\odot}$), then nucleosynthetic signatures of 
pair-instability supernovae should be found in the abundance patterns 
of extremely metal poor stars \cite{heger02,tumlinson04}. The absence 
of such patterns to date has been interpreted as evidence that the 
Population III IMF lacks very massive stars \cite{tumlinson06}.

\begin{figure}\label{fig:emp}
  \includegraphics[height=.26\textheight]{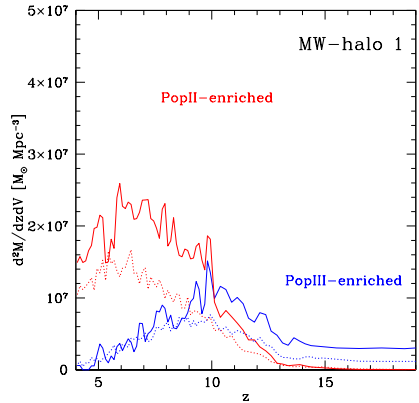}
  \caption{Formation rate per unit redshift of extremely metal poor
    gas ($Z \sim 10^{-3.5} Z_{\odot}$) as measured from a
   cosmological simulations that take into account 
   radiative feedback for Population III formation, self-enrichment of 
   halos, and metal winds propagating at $60~\mathrm{km~s^{-1}}$ (see 
   \citealt{trenti09b}). The rate of gas enriched by Pop~III stars is shown as a
   blue lines while the rate for Pop~II-enriched gas is shown as red
   lines. Solid lines refer to a Milky Way like halo. Dotted lines to an
   average region of the Universe. The majority of extremely low
   metallicity gas is enriched at relatively low redshift ($z \lesssim
   10$) by wind outflows. Figure from \cite{trenti_shull10}.}
\end{figure}

However, this conclusion ignores wind enrichment and hence rests on the
assumption that observed extremely metal poor stars ($Z \sim 10^{-3.5}
Z_{\odot}$) formed out of gas that was enriched by Population III stars. 
Our cosmological simulations instead show that metal outflows from dwarf 
galaxies enrich the majority of the extremely metal poor gas present in 
Milky Way progenitors (Figure~\ref{fig:emp} and \cite{trenti_shull10}). 
This implies that galactic archeology primarily probes the IMF of 
Population II stars, so the absence of pair-instability signatures is 
not surprising. The IMF of Population III stars remains an open question 
for the decade to come.


\begin{theacknowledgments} It is a pleasure to thank Daniel Whalen,
  Naoki Yoshida and Volker Bromm for the excellent organization of the meeting. MT acknowledges support from NASA (NNX07AG77G) and NSF (AST07-07474). \end{theacknowledgments}



\bibliographystyle{aipproc}   

\begin{thebibliography}{29}
\expandafter\ifx\csname natexlab\endcsname\relax\def\natexlab#1{#1}\fi
\providecommand{\enquote}[1]{``#1''}
\expandafter\ifx\csname url\endcsname\relax
  \def\url#1{\texttt{#1}}\fi
\expandafter\ifx\csname urlprefix\endcsname\relax\def\urlprefix{URL }\fi
\providecommand{\eprint}[2][]{\url{#2}}

\bibitem[{Tegmark} et~al.(1997)]{tegmark97}
M.~{Tegmark}, J.~{Silk}, M.~J. {Rees}, A.~{Blanchard}, T.~{Abel}, and
  F.~{Palla}, \emph{\apj} \textbf{474}, 1 (1997).

\bibitem[{Abel} et~al.(2002)]{abel02}
T.~{Abel}, G.~L. {Bryan}, and M.~L. {Norman}, \emph{Science} \textbf{295},
  93  (2002).

\bibitem[{Bromm} and {Larson}(2004)]{bromm04}
V.~{Bromm}, and R.~B. {Larson}, \emph{\araa} \textbf{42}, 79 (2004).

\bibitem[{O'Shea} and {Norman}(2007)]{oshea07}
B.~W. {O'Shea}, and M.~L. {Norman}, \emph{\apj} \textbf{654}, 66 (2007).


\bibitem[{Turk} et~al.(2009)]{turk09}
M.~J. {Turk}, T.~{Abel}, and B.~{O'Shea}, \emph{Science} \textbf{325}, 601
  (2009).

\bibitem[{Naoz} et~al.(2006)]{naoz06}
S.~{Naoz}, S.~{Noter}, and R.~{Barkana}, \emph{\mnras} \textbf{373}, L98
  (2006).

\bibitem[{Trenti} and {Stiavelli}(2007)]{ts07a}
M.~{Trenti}, and M.~{Stiavelli}, \emph{\apj} \textbf{667}, 38 (2007).

\bibitem[{Heger} et~al.(2003)]{heger03}
A.~{Heger}, C.~L. {Fryer}, S.~E. {Woosley}, N.~{Langer}, and D.~H. {Hartmann},
  \emph{\apj} \textbf{591}, 288 (2003).

\bibitem[{Trenti} et~al.(2008)]{tss08}
M.~{Trenti}, M.~R. {Santos}, and M.~{Stiavelli}, \emph{\apj} \textbf{687}, 1
  (2008).

\bibitem[{Haiman} et~al.(1997)]{haiman97}
Z.~{Haiman}, M.~J. {Rees}, and A.~{Loeb}, \emph{\apj} \textbf{476}, 458
  (1997).

\bibitem[{Trenti} and {Stiavelli}(2009)]{ts09}
M.~{Trenti}, and M.~{Stiavelli}, \emph{\apj} \textbf{694}, 879 (2009).

\bibitem[{Ricotti} et~al.(2008)]{ricotti08}
M.~{Ricotti}, N.~Y. {Gnedin}, and J.~M. {Shull}, \emph{\apj} \textbf{685},
  21 (2008).

\bibitem[{Trenti} et~al.(2009)]{trenti09b}
M.~{Trenti}, M.~{Stiavelli}, and J.~M. {Shull}, \emph{\apj} \textbf{700},
  1672  (2009).

\bibitem[{Stiavelli} and {Trenti}(2010)]{stiavelli09}
M.~{Stiavelli}, and M.~{Trenti},  \emph{\apjl}
  \textbf{716}, L190 (2010).

\bibitem[{Tornatore} et~al.(2007)]{tornatore07}
L.~{Tornatore}, A.~{Ferrara}, and R.~{Schneider}, \emph{\mnras} \textbf{382},
  945 (2007).

\bibitem[{Bagla} and {Ray}(2005)]{bagla05}
J.~S. {Bagla}, and S.~{Ray}, \emph{\mnras} \textbf{358}, 1076 (2005).
 

\bibitem[{Trenti} et~al.(2010{\natexlab{a}})]{trenti10a}
M.~{Trenti}, B.~D. {Smith}, E.~J. {Hallman}, S.~W. {Skillman}, and J.~M.
  {Shull}, \emph{\apj} \textbf{711}, 1198 (2010{\natexlab{a}}).
 

\bibitem[{Yoshida} et~al.(2007)]{yoshida07}
N.~{Yoshida}, K.~{Omukai}, and L.~{Hernquist}, \emph{\apjl} \textbf{667},
  L117 (2007).

\bibitem[{Smith} et~al.(2009)]{smith_b09}
B.~D. {Smith}, M.~J. {Turk}, S.~{Sigurdsson}, B.~W. {O'Shea}, and M.~L.
  {Norman}, \emph{\apj} \textbf{691}, 441 (2009).

\bibitem[{Schneider} and {Omukai}(2010)]{schneider10}
R.~{Schneider}, and K.~{Omukai}, \emph{\mnras} \textbf{402}, 429 (2010).

\bibitem[{Tan} and {McKee}(2008)]{tan08}
J.~C. {Tan}, and C.~F. {McKee},  2008, vol. 990 of \emph{AIP Conference
  Series}, pp. 47.

\bibitem[{Trenti} et~al.(2010{\natexlab{b}})]{trenti10b}
M.~{Trenti}, et~al., \emph{\apjl}
  \textbf{714}, L202 (2010{\natexlab{b}}).

\bibitem[{Freeman} and {Bland-Hawthorn}(2002)]{freeman02}
K.~{Freeman}, and J.~{Bland-Hawthorn}, \emph{\araa} \textbf{40}, 487
  (2002).

\bibitem[{Beers} and {Christlieb}(2005)]{beers05}
T.~C. {Beers}, and N.~{Christlieb}, \emph{\araa} \textbf{43}, 531 (2005).

\bibitem[{Frebel} et~al.(2007)]{frebel07}
A.~{Frebel}, J.~L. {Johnson}, and V.~{Bromm}, \emph{\mnras} \textbf{380},
  L40  (2007).

\bibitem[{Tumlinson} et~al.(2004)]{tumlinson04}
J.~{Tumlinson}, A.~{Venkatesan}, and J.~M. {Shull}, \emph{\apj} \textbf{612},
  602  (2004).

\bibitem[{Heger} and {Woosley}(2002)]{heger02}
A.~{Heger}, and S.~E. {Woosley}, \emph{\apj} \textbf{567}, 532 (2002).


\bibitem[{Tumlinson}(2006)]{tumlinson06}
J.~{Tumlinson}, \emph{\apj} \textbf{641}, 1  (2006).


\bibitem[{Trenti} and {Shull}(2010)]{trenti_shull10}
M.~{Trenti}, and J.~M. {Shull}, \emph{\apj} \textbf{712}, 435  (2010).

\end{thebibliography}

\end{document}